\documentclass[epjCONF]{svjour}
\usepackage{graphics}
\usepackage[varg]{txfonts} % Times fonts
\usepackage[latin1]{inputenc}
\usepackage{amstext}
\usepackage[usenames,dvipsnames]{color}

\session-title{Conference Title, to be filled}
\begin{document}
\title{Constraining angular momentum transport processes in stellar interiors with red-giant stars in the open cluster NGC6819}
\author{Lagarde, N\inst{1}\fnmsep\thanks{\email{lagarde@bison.ph.bham.ac.uk}} \and Miglio, A.\inst{1,2} \and Eggenberger, P.\inst{3} \and Montalb\'an, J\inst{4} \and Bossini, D\inst{1,2}}
\institute{School of Physics and Astronomy, University of Birmingham, Edgbaston, Birmingham B15 2TT, UK \and Stellar Astrophysics Centre (SAC), Department of Physics and Astronomy, Aarhus University, Ny Munkegade 120, 8000 Aarhus C, Denmark \and Geneva Observatory, University of Geneva, Chemin des Maillettes 51, 1290 Versoix, Switzerland \and Dipartimento di Astronomia, Universit{\`a} di Padova, Padova, Italy}
\abstract{Clusters are excellent test benches for verification and improvement of stellar evolution theory. The recent detection of solar-like oscillations in G-K giants in the open cluster NGC6819 with \textit{Kepler} provides us with independent constraints on the masses and radii of stars on the red giant branch, as well as on the distance to clusters and their ages. We present, for NGC6819, evolutionary models by considering rotation-induced mixing ; and the theoretical low-l frequencies of our stellar models.
} %end of abstract
\maketitle
\section{Introduction}
\label{intro}

In recent years, a large number of asteroseismic data were obtained for different kinds of stars by space missions (\cite{DeRidder09},\cite{Bedding10}), allowing the detection and characterization of solar-like oscillations in a large number of red giants. \textit{Kepler} data of giants in the open cluster NGC6819 represent an unique opportunity to test stellar evolution models, allowing access to the properties of the stellar interiors of red giant stars, and giving us information on the internal rotation of giant stars (\cite{Deheuvels12},\cite{Deheuvels14}, \cite{Beck12},\cite{Mosser12}).
As underlined by  \cite{Eggenberger12} in the case of the giant star KIC8366239, a discrepancy exists between the rotation profile deduced from asteroseismic observations and the profiles predicted from models including shellular rotation and related meridional flows and turbulence. They show that a most efficient mechanism must in action to extract angular momentum from the core of this star, and other red giant stars.

\section{Models }

We use the code STAREVOL (e.g. \cite{Lagarde12}) to compute stellar evolution models taking into account rotation-induced processes following the formalism by \cite{Zahn92} and \cite{MaeZah98}, and thermohaline mixing as described by \cite{ChaZah07a}. 
As proposed by \cite{Eggenberger12} models include an additional viscosity, corresponding to an physical process which can transport the angular momentum in radiative zone (see Eq.1 of \cite{Lagarde14}). This corresponds well to the two main mechanisms currently proposed to efficiently extract angular momentum from the central core of a solar-type star having a strong impact on the transport of angular momentum. Frequencies have been computed with the Liege Oscillation code (\cite{Scuflaire08}).

%\begin{equation}
%\underbrace{\rho\frac{d(r^{2} \Omega)}{d t}}_\text{\parbox{4cm}{\centering stellar \\[-4pt] contraction/expansion}}=\underbrace{\frac{1}{5r^{2}}\frac{\partial}{\partial r}(\rho r^{4} \Omega U_{r})}_\text{\parbox{4cm}{\centering advection of \\[-4pt] angular momentum\\[-4pt] by meridional circulation}} + \underbrace{\frac{1}{r^2}\frac{\partial}{\partial r}\left(r^{4}\rho (D{\color{red}{+\nu_{add}}})\frac{\partial \Omega}{\partial r}\right)}_\text{\parbox{4cm}{\centering diffusion effect of \\[-4pt] shear-induced turbulence\\[-4pt] and additional viscosity}},
% \label{eqAM}
 %\end{equation}

\begin{figure}
\centering{
\resizebox{0.31\columnwidth}{!}{%
  \includegraphics{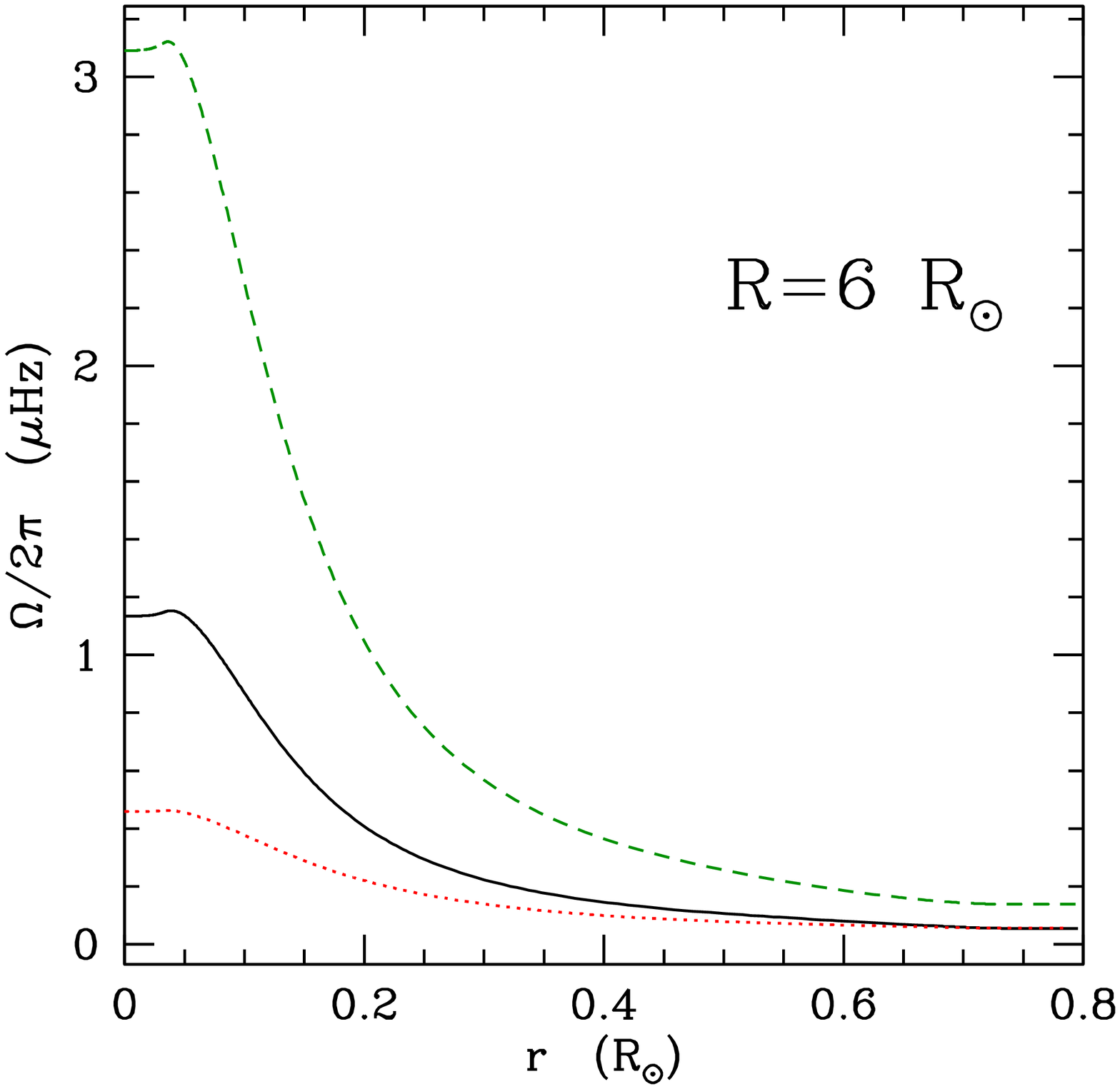}}
\resizebox{0.31\columnwidth}{!}{%
  \includegraphics{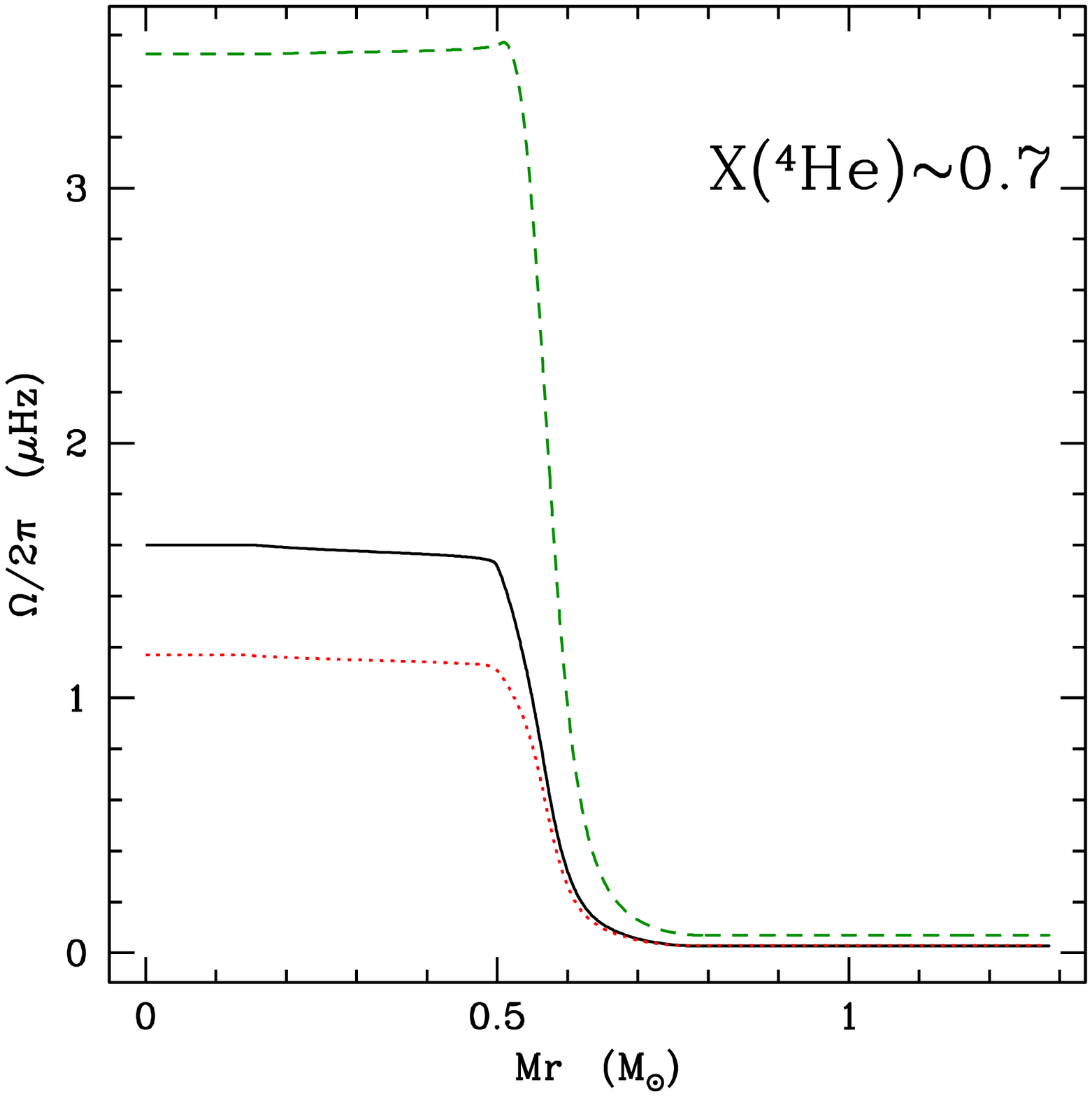} }\\
\resizebox{0.335\columnwidth}{!}{%
  \includegraphics{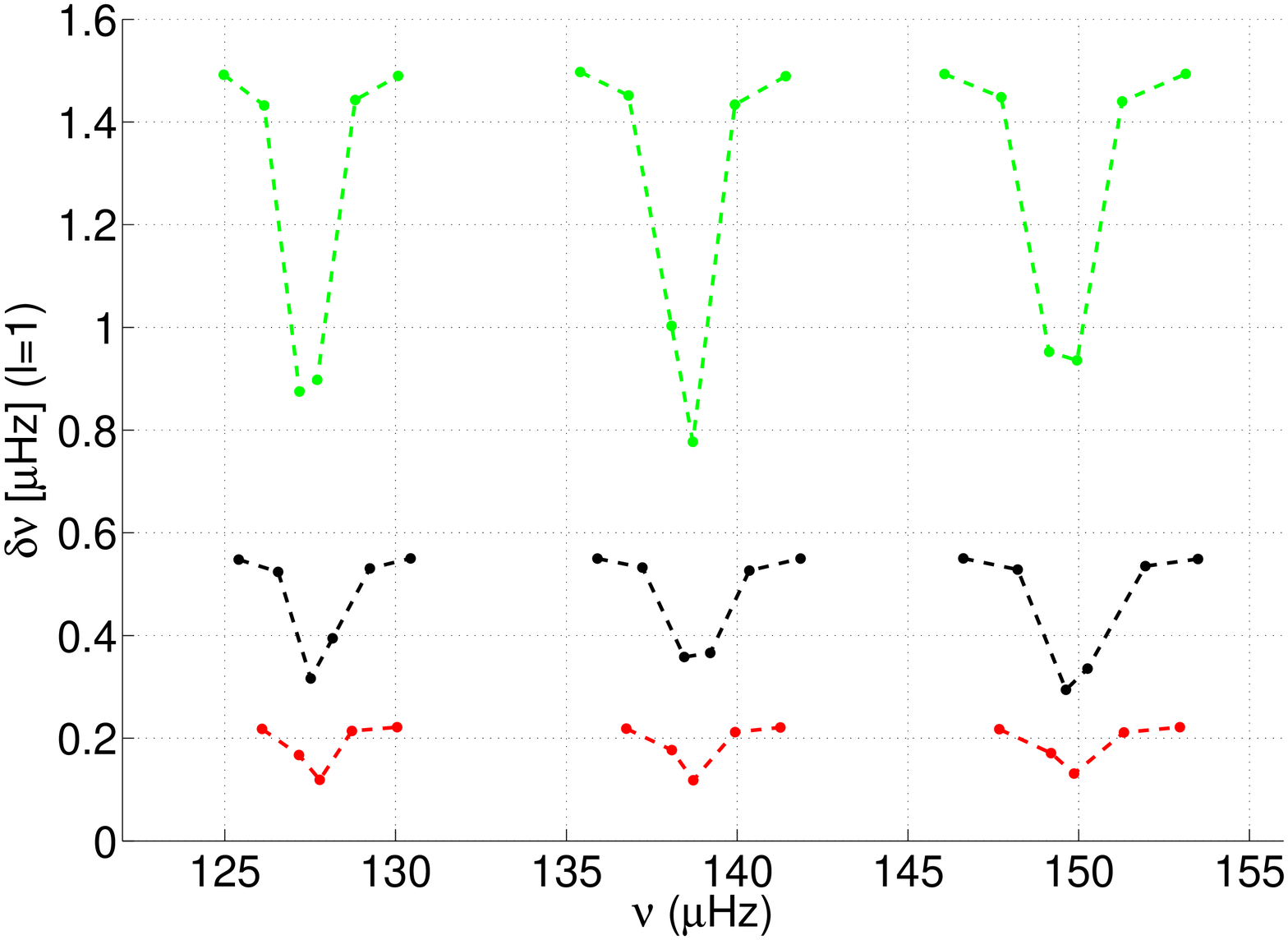} }
\resizebox{0.31\columnwidth}{!}{%
  \includegraphics{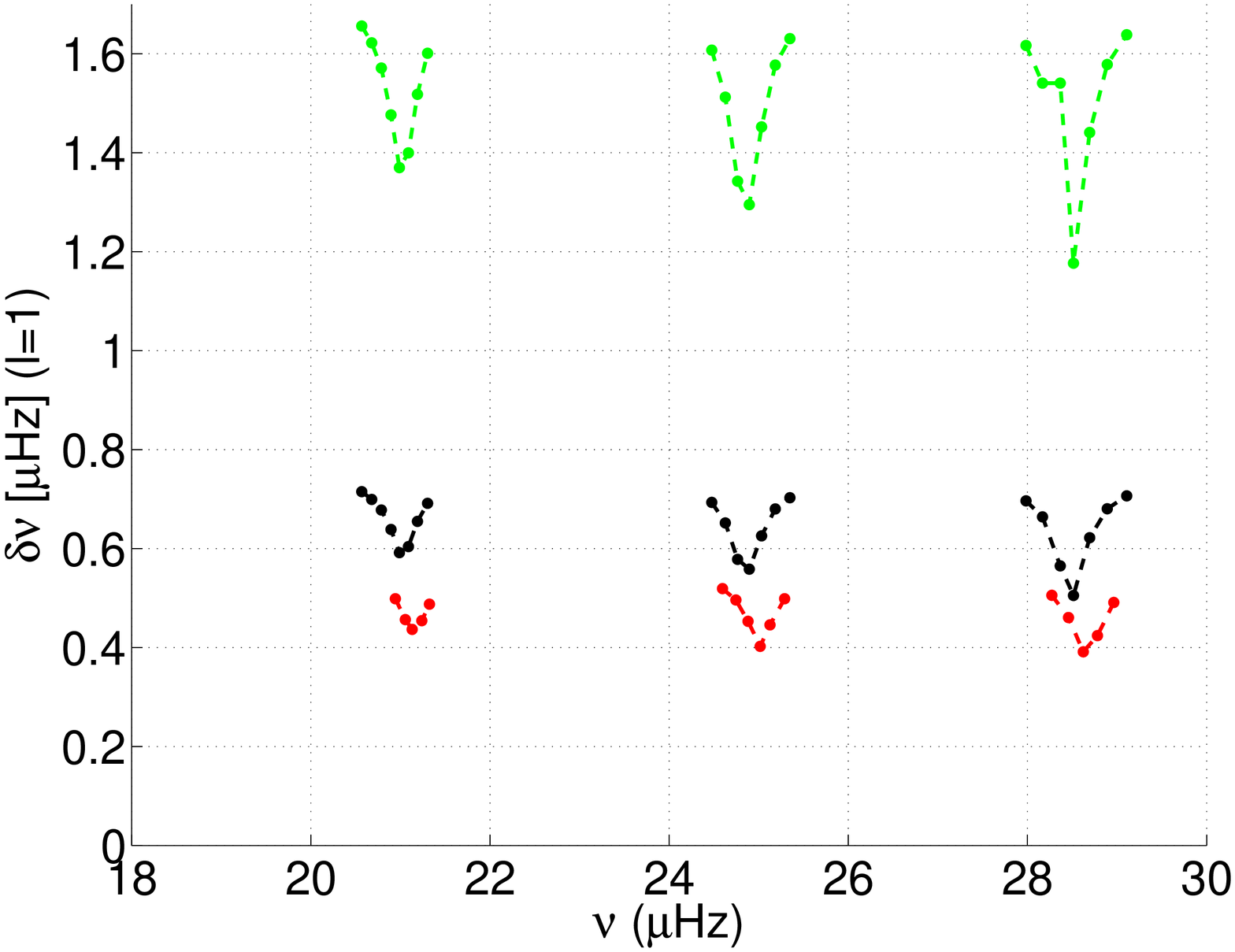} }}
\caption{Top panels: Rotation profiles during the RGB at R=6R$_{\odot}$ (left) and central He-burning phase at X($^4$He)$\sim$0.7 (right) for models including an additional viscosity and at different initial velocity at the ZAMS: V$_{\rm{ZAMS}}$=50 km.s$^{-1}$ and $\nu_{add}$=3.10$^{4}$cm$^{2}$.s$^{-1}$ (green-dashed line), V$_{\rm{ZAMS}}$=20 km.s$^{-1}$ and $\nu_{add}$=3.10$^{4}$cm$^{2}$.s$^{-1}$ (black-solid line),V$_{\rm{ZAMS}}$=20 km.s$^{-1}$ and $\nu_{add}$=5.10$^{4}$cm$^{2}$.s$^{-1}$ (red-dotted line). Bottom panels: Rotational splittings of l=1 modes for the same models, during the RGB (left), and the clump (right).}
\label{fig:1}       % Give a unique label
\end{figure}

Figure \ref{fig:1} displays the effects of additional viscosity and different initial rotation velocity on the theoretical rotational profiles and splittings for a 1.5M$_{\odot}$ model at solar metallicity for a given stellar radius on the red giant branch and a given central mass fraction of helium in the clump (core-He burning phase).
We show that an increase in $\nu_{\rm{add}}$ results in a more efficient transport of angular momentum, hence a flatter rotation profile in the radiative zone (RGB, top-left), and a lower core rotation rate in the He core (clump, top-right). 
As in these models rotational splittings ($\delta \nu$) are sensitive to the the rotation rate of the stellar core, $\delta \nu$ decreases when $\nu_{\rm{add}}$ increases (bottom panels). Bottom panels of Fig. \ref{fig:1} show that the ratio of the splittings of modes in the wings (more gravity-like modes) to those for modes at the center (more pressure-like modes) is lower in clump than in RGB models.
$\delta \nu$ values allow us then to estimate the efficiency ($\nu_{\rm{add}}$) of the  additional physical process working during the RGB.

% $\nu_{\rm{add}}$ allows us to determine the efficiency of an additional physical process needed on the RGB, which can be constrained thanks to asteroseismic measurements.
 
\begin{thebibliography}{}
\bibitem{DeRidder09}
De Ridder, J., Barban, C., Baudin, F., et al., Nature, \textbf{459}, (2009) 398
\bibitem{Bedding10}
Bedding, T. R., Huber, D., Stello, D., et al., ApJ, \textbf{713}, (2010) L176
\bibitem{Baglin06}
Baglin, A., Auvergne, M., et al., in 36th COSPAR Scientific Assembly, \textbf{Vol. 36}, (2006) 3749
\bibitem{Borucki10}
Borucki, W. J., Koch, D., Basri, G., et al., Science, \textbf{327}, (2010) 977
\bibitem{Deheuvels14}
Deheuvels, S., Do{\v g}an, G., Goupil, M. J., et al., A\&A, \textbf{564}, (2014) A27 
\bibitem{Deheuvels12}
Deheuvels, S., Garc{\'i}a, R. A., Chaplin, W. J., et al., ApJ, \textbf{756}, (2012) 19
\bibitem{Beck12}
Beck, P. G., Montalban, J., Kallinger, T., et al., Nature, \textbf{481}, (2012) 55
\bibitem{Mosser12}
Mosser, B., Goupil, M. J., Belkacem, K., et al., A\&A, \textbf{548}, (2012) A10
\bibitem{Eggenberger12}
Eggenberger, P., Montalb{\'a}n, J., \& Miglio, A., A\&A, \textbf{544},  (2012) L4
\bibitem{Lagarde12}
Lagarde, N., Decressin, T., Charbonnel, C., Eggenberger, P., et al., A\&A, \textbf{543}, (2012), A108
\bibitem{Lagarde14}
Lagarde, N., Eggenberger P., Miglio, A. and Montalb{\'a}n, J., SF2A-2014, (2014)
\bibitem{Zahn92}
Zahn, J.-P., A\&A, \textbf{265}, (1992) 115
\bibitem{MaeZah98}
Maeder, A. \& Zahn, J.-P., A\&A, \textbf{334}, (1998) 1000
\bibitem{ChaZah07a}
Charbonnel, C. \& Zahn, J.-P., A\&A, \textbf{476}, (2007) L29
\bibitem{Scuflaire08}
Scuflaire et al, ASS, \textbf{316}, (2008) 149
\end{thebibliography}
\end{document}